\documentclass[preprintnumbers,amsmath,amssymb,floatfix,10pt,prd,twocolumn,superscriptaddress,nofootinbib]{revtex4-2}
\usepackage{bm}
\usepackage{amsfonts}
\usepackage{latexsym}
\usepackage[latin1]{inputenc}
\usepackage{graphicx}
\usepackage{amsmath}
\usepackage{palatino}
\usepackage{mathpazo}
\usepackage{textcomp}
\usepackage{textcomp}
\usepackage{mathrsfs}
\usepackage{float}
\usepackage{booktabs}
\usepackage{dcolumn}
\usepackage{ragged2e}
\usepackage{hyperref}
\usepackage{tensor}
\usepackage[version=3]{mhchem}
\hypersetup{colorlinks,citecolor=blue}
\hypersetup{colorlinks=true,linkcolor=red,filecolor=magenta,    urlcolor=blue}
\usepackage{amsmath}
\usepackage{xcolor}
\usepackage{enumitem}
\usepackage{orcidlink}
\usepackage{epsfig}
\usepackage{subfigure}
\usepackage{commath}
\usepackage{caption}
\usepackage{bigints}
\begin{document}

\title{Inflationary Dynamics and CMB Constraints in $f(T,L_m)$ gravity}
\author{Suchita Patel\orcidlink{0009-0003-9597-7438}}
\email{kmsuchitapatel017@gmail.com}
\affiliation{Department of Mathematics, Birla Institute of Technology and Science, Pilani, Hyderabad Campus, Jawahar Nagar, Kapra Mandal, Medchal District, Telangana 500078, India}%
\author{Sai Swagat Mishra\orcidlink{0000-0003-0580-0798}}
\email{saiswagat009@gmail.com}
\affiliation{Department of Astronomy, School of Physical Sciences, University of Science and Technology of China, Hefei, Anhui
230026, China}
\affiliation{School of Astronomy and Space Science, University of Science and Technology of China, Hefei, Anhui 230026, China}%

\author{P.K. Sahoo\orcidlink{0000-0003-2130-8832}}
\email{pksahoo@hyderabad.bits-pilani.ac.in}
\affiliation{Department of Mathematics, Birla Institute of Technology and Science, Pilani, Hyderabad Campus, Jawahar Nagar, Kapra Mandal, Medchal District, Telangana 500078, India}
\begin{abstract}
We investigate slow-roll inflation in a modified teleparallel framework with a non-minimal coupling between torsion and the matter sector. Focusing on Starobinsky-type plateau inflation, we derive the modified background dynamics and the corresponding slow-roll observables, and examine how the gravitational couplings affect the scalar spectral index, tensor-to-scalar ratio, and running of the spectral index. The model continuously recovers the standard Starobinsky scenario in the appropriate weak-coupling limit, while the matter--geometry coupling produces a non-monotonic modification of the inflationary predictions. We find that weak couplings leave the standard predictions essentially unchanged, whereas stronger couplings can significantly shift the scalar spectral index and eventually become incompatible with current cosmic microwave background constraints. The quadratic torsion correction has a comparatively mild impact within the controlled low-energy regime, primarily affecting the tensor amplitude. We confront the resulting predictions with the Planck and BICEP/Keck constraints and find that the model remains viable over a broad range of the inflationary parameter space. Our results demonstrate that the matter--geometry coupling can leave a potentially testable imprint on primordial observables while retaining the successful plateau-inflation phenomenology of the Starobinsky scenario.

\noindent\textbf{Keywords:} Inflation, torsion--matter coupling, Starobinsky potential, cosmic microwave background
\end{abstract}

\maketitle

\section{Introduction}
Based on General Relativity (GR) \cite{Einstein:1916} and the Lambda Cold Dark Matter ($\Lambda$CDM) paradigm, the conventional cosmological framework has been very successful in explaining the Universe's evolution. Numerous observational data, including measurements of temperature and polarization anisotropies in the cosmic microwave background (CMB) \cite{Planck:2018VI, Planck:2018X, WMAP:2013, BICEP:2021xfz, SPT-3G:2014dbx, AtacamaCosmologyTelescope:2025nti} are in great agreement with its theoretical predictions. However, there are still several basic conceptual issues with this framework. The horizon and flatness issues are among the best known, and they pose a basic question: how can the observable Universe show such a high degree of homogeneity across regions that would never have been able to establish causal contact under the assumption of a solely decelerating expansion?

Cosmic inflation provides an elegant solution to both of these puzzles. Inflation drives the spatial curvature toward flatness and expands a small, causally connected region to include the entire observable Universe today by introducing a brief period of accelerated expansion in the early Universe \cite{Starobinsky:1980te, Guth:1980zm, Linde:1981mu, Albrecht:1982wi}. In addition to resolving these initial-condition issues, inflation produces the seeds of cosmic structure, which is perhaps even more amazing. The density disturbances we subsequently see as temperature anisotropies in the CMB are the result of quantum fluctuations in the field driving inflation being stretched to cosmological scales \cite{Hawking:1982cz, Guth:1982ec, Bardeen:1983qw}.

The simplest and extensively studied realization of this framework is single-field slow-roll inflation, in which a scalar field, the inflaton $\phi$, slowly rolls down a potential $V(\phi)$  \cite{Linde:1983gd}. Here, ``slowly" has a definite dynamical meaning: the kinetic energy of the field must be tiny compared to its potential energy for a long time. This condition is quantified in terms of the slow-roll parameters \cite{Liddle:1992wi, Liddle:1994dx, Lyth:1998xn}. This simple framework allows for a wide variety of possible forms: power-law potentials, hilltop potentials, natural inflation with periodic potentials, the Starobinsky $R^2$ model, and many others, each making distinct predictions for the scalar spectral index $n_s$ and tensor-to-scalar ratio $r$  \cite{Martin:2013tda, Planck:2018X, Linde:1983gd, Freese:1990rb, Boubekeur:2005zm}. Accurate measurements from Planck and BICEP/Keck have been pivotal, eliminating ideas like chaotic quartic inflation that previously seemed plausible \cite{Planck:2018X, BICEP:2021xfz}. The relationship between theory and data is, in various aspects, the hallmark of current inflationary model development: a potential is regarded as useful only if it produces accurate forecasts in the $(n_s, r)$ domain.

GR remains the best description of gravity at the scales we can test it \cite{Will:2014kxa}. Dark matter and dark energy govern the late-time evolution of the Universe, and GR does not predict either of them from first principles \cite{SupernovaSearchTeam:1998fmf, SupernovaCosmologyProject:1998vns, Peebles:2002gy}. This discrepancy has motivated a long-standing search for alternative theories of gravity \cite{Clifton:2011jh, Nojiri:2017ncd, Capozziello:2011et, Sotiriou:2008rp}. These modified theories are widely used in early-Universe cosmology with the hope that their predictions are closer to inflationary data \cite{Bamba:2012cp, DeFelice:2010aj, Odintsov:2026fyc}.

One way to a theory more general than GR is to go beyond Riemannian geometry. In GR, the gravitational interaction is carried only by curvature, with zero torsion and non-metricity. There are two alternatives, one of them is the teleparallel equivalent of GR (TEGR), constructed based on the Weitzenb\"ock connection, which has zero curvature and zero non-metricity, but a non-vanishing torsion \cite{Sauer:2004hj, Hayashi:1979qx, Aldrovandi:2013wha}. Here, the dynamical variable is the tetrad field instead of the metric, and the gravitational Lagrangian is the torsion scalar $T$. Secondly, symmetric teleparallel gravity, which takes the complementary route. Zero curvature and zero torsion are present while gravity is carried by the non-metricity tensor \cite{Nester:1998mp, BeltranJimenez:2017tkd}. At the level of the field equations, the three descriptions are dynamically equivalent, but their extensions are not.

The teleparallel gravity has been generalized to $f(T)$ gravity where the Lagrangian is an arbitrary function of the torsion scalar \cite{Ferraro:2006jd, Bengochea:2008gz, Linder:2010py, Mishra:2025rhi}. The field equations of  $f(T)$ gravity are second order, unlike those of $f(R)$ gravity, which are fourth order \cite{Ferraro:2006jd, Cai:2015emx}. The framework permits self-accelerating solutions without a cosmological constant, and its cosmological and astrophysical implications have been examined in depth \cite{Cai:2015emx, Bamba:2010wb, ElHanafy:2014tzj, Krssak:2018ywd, Kavya:2024ssu}. 
$f(T)$ gravity has also been applied to the inflationary period, where the updated Friedmann equations modify the slow-roll parameters and shift the anticipated $n_s$ and $r$ \cite{Ferraro:2006jd, Rezazadeh:2015dza, Bamba:2016wjm}.

Allowing for an explicit coupling between geometry and matter is a logical progression from $f(T)$ gravity. The gravitational sector is directly dependent on the matter fields in theories of this class, which do not rely on the minimal coupling assumption. There are two options for matter variables. The first is the trace $\mathcal{T}$ of the energy-momentum tensor, which on the torsion side produces $ f(T,\mathcal{T})$ gravity \cite{Harko:2014trace, Mishra:2023onl, Mishra:2024oln}. This was preceded on the torsion side by a nonminimal coupling of a restricted form \cite{Harko:2014sja}. The second is the Lagrangian density of the matter, defined as $\mathcal{L}_m$. The gravity of  $f(R,\mathcal{L}_m )$ was used to open this route in the curvature setting \cite{Harko:2010mv}. This gravity was then carried to the non-metricity setting by  $f(Q,\mathcal{L}_m)$ \cite{Hazarika:2024alm, Rana:2025qjw}. The pattern is completed by its torsion counterpart, $f(T,\mathcal{L}_m)$  gravity, first introduced by a specific form of coupling by \cite{Harko:2014sja} and then in a more general form by \cite{SwagatMishra:2025scq, Cruz:2025fuk}, which is the framework we employ here. The trace is a single scalar, whereas $\mathcal{L}_m$
ensures that the matter fields retain their full dynamical content, which makes it more general.

The repercussions of this coupling are structural rather than incidental. The field equations acquire terms that are proportional to the derivatives of $f$ with respect to $\mathcal{L}_m$,
which, depending on the matter content itself, function as an effective gravitational connection. The energy-momentum tensor is not covariantly conserved in general, and energy may be exchanged between the gravitational and matter sectors as a result \cite{Cruz:2025fuk}. The curvature and non-metricity versions of the theory are already familiar with the same feature \cite{Harko:2010mv, Hazarika:2024alm}. It provides a framework for late-time acceleration that does not need a cosmological constant. Early-to-late transition, gravitational baryogenesis, Chaplygin gas, and big bang nucleosynthesis are some of the topics that have been studied recently in order to map out these effects \cite{SwagatMishra:2025scq, Cruz:2025sat, Cruz:2025fuk, Samaddar:2026ter}.

In this work, we study cosmological inflation within the $f(T,\mathcal{L}_m)$ framework. This paper is organized as follows. In Sec \ref{sec:background}, we will give a brief introduction to the mathematical formulation in the $f(T,\mathcal{L}_m)$ framework, and then we review the slow-roll inflationary dynamics in standard GR and set out the equations that govern the scalar field and the inflationary parameters. Sec \ref{sec:slowroll} derives the inflationary dynamics in the present framework through our chosen model and obtains the observables for the Starobinsky potential. Sec \ref{sec:interpretaion} interprets these results. Finally, Sec \ref{sec:conclsn} summarizes the key results and discusses what they imply for the viability of $f(T,\mathcal{L}_m)$ gravity as a framework for early Universe inflation.

\section{\(f(T,\mathcal{L}_m)\) Gravity Framework}\label{sec:background} 

This section establishes the theoretical basis of \( f(T, \mathcal{L}_m) \) gravity and methodically derives the core field equations that describe the evolution of the cosmological background. In the teleparallel picture, the Levi-Civita connection used in GR is replaced by the Weitzenb\"ock connection, so that gravity is encoded in torsion rather than in spacetime curvature. In this formulation, the tetrad fields \( e^{\mu}_{\ i} \) take the place of the metric tensor as the primary dynamical variables. These tetrads carry the gravitational degrees of freedom and obey an orthonormality condition, which preserves the consistency and geometric structure of the spacetime manifold
\begin{align}\label{tetrad}
    e^{\ \alpha}_{i}e^{\ \beta}_{j}g_{\alpha\beta}=\eta_{ij},
\end{align}
where  $g_{\alpha\beta}$ is the spacetime metric and $\eta_{ij}$ is the Minkowski metric of the tangent space given as  $\eta_{ij}=diag(1,-1,-1,-1)$.

Since $e^{\alpha}_{\ \ i} e_{\beta}^{\ \ i}=\delta^{\alpha}_{\ \ \beta}$, therefore equation (\ref{tetrad}) can also be written as 
\begin{align}
    e^{i}_{\  \alpha}e^j_{\  \beta}\eta_{ij}=g_{\alpha\beta}.
\end{align}

The Weitzenb\"{o}ck's connection \cite{Aldrovandi:2013wha} in terms of tetrads is defined as
\begin{align}\label{weigenbock connection}
    \tilde{\Gamma}^{\rho}_{\alpha\beta}=e_i^{\ \rho}\partial_{\beta}e^{\ \ i}_{\alpha}.
\end{align}

Taking the antisymmetric component of the connection \eqref{weigenbock connection} leads to the construction of the following torsion tensor,
\begin{align}\label{Torsion}
    T^{\lambda}_{\ \ \alpha\beta}=\tilde{\Gamma}^{\lambda}_{\beta\alpha}-\tilde{\Gamma}^{\lambda}_{\alpha\beta}=e_i^{\ \ \lambda}(\partial_{\alpha}e^i_{\ \ \beta}-\partial_{\beta}e^i_{\ \ \alpha}).
\end{align}

The contorsion tensor is defined as
\begin{align}
    K^{\alpha\beta}_{ \ \  \  \  \lambda}=-\frac{1}{2}(T^{\alpha\beta}_{\ \  \  \  \lambda}-T^{\beta\alpha}_{\ \  \  \  \lambda}-T_{\lambda}^{\ \ \alpha\beta}).
\end{align}

The torsion scalar is derived from the following contraction
\begin{align}\label{torsion scalar}
    T= T^{\lambda}_{\  \ \alpha\beta}S_{\lambda}^{\ \  \alpha\beta}.
\end{align}
where $ S_{\lambda}^{\ \  \alpha\beta}$ is called the superpotential tensor given as 
\begin{align}
    S_{\lambda}^{\ \  \alpha\beta}=\frac{1}{2}(K^{\alpha\beta}_{ \ \  \  \  \lambda}+\delta_{\lambda}^{\alpha}T^{\mu\beta}_{\ \ \mu}-\delta_{\lambda}^{\beta}T^{\mu\alpha}_{\ \ \mu}).
\end{align}

We now define the action for the $f(T, \mathcal{L}_m)$ gravity theory as follows \cite{SwagatMishra:2025scq}
\begin{align}\label{action}
    \mathcal{S}=\frac{1}{2\kappa}\int dx^4 e(T+f(T,\mathcal{L}_m))+\int dx^4 e\mathcal{L}_m .
\end{align}

Note that $e=det(e^{i}_{\ \alpha})$. We take the matter Lagrangian to depend only on the tetrad fields $e^i_{\ \alpha}$, without involving any of their derivatives. Under this assumption, performing a variation of the action with respect to the tetrads directly leads to the associated field equations,
\begin{align}
        &(1+f_T)\left[e^{-1}\partial_\alpha(ee_i^{\ \lambda}S_\lambda^{\ \alpha\beta})-e_i^{\ \rho}T^{\lambda}_{\ \alpha\rho}S_\lambda^{\ \beta\alpha}\right] \\ \nonumber 
       &+e_i^{\ \lambda}S_\lambda^{\ \alpha\beta}(f_{TT}\partial_\alpha T+f_{T\mathcal{L}_m}\partial_\alpha \mathcal{L}_m)-\frac{f_{\mathcal{L}_m}}{4}\left(e_i^{\ \lambda}\overset{em}{T}{}_\lambda^{\ \beta}+e_i^{\ \lambda}\mathcal{L}_m\right)\\ \nonumber 
       &+e_i^{\ \beta}\left(\frac{T+f}{4}\right)=4\pi G e_i^{\ \lambda}\overset{em}{T}{}_\lambda^{\ \beta}\,,
    \label{field equations}
\end{align}
where $\overset{em}{T}{}_\lambda^{\ \beta}$ represents the energy-momentum tensor of a perfect fluid \cite{Aldrovandi:2013wha}
\begin{align} 
 \frac{\delta e\mathcal{L}_m}{\delta e^i_{\ \beta}}=-ee_i^{\ \lambda}T_\lambda^{\ \beta},
 \end{align} 
 and $f_T=\frac{\partial f}{\partial T}$, $f_{TT}=\frac{\partial^2 f}{\partial T^2}$, $f_{T{\mathcal{L}_m}}=\frac{\partial^2 f}{\partial T\partial{\mathcal{L}_m}}$, $f_{\mathcal{L}_m}=\frac{\partial f}{\partial{\mathcal{L}_m}}$.
 
We adopt the spatially flat FLRW metric, considering the spatial isotropy and homogeneity of the universe
\begin{align}\label{FLRW}
 ds^2=dt^2-a^2(t)\,\delta_{ij}\,dx^i dx^j,
\end{align}
 where $a(t)$ is the scale factor as a function of time.
 Consequently, the suitable choice of tetrads for the metric (\ref{FLRW}) is the following diagonal form,
 \begin{align}\label{tetrad diagonal form}
    e_i^{\ \alpha}=diag(1,a(t),a(t),a(t)).
 \end{align}
 
 Substituting the tetrads from equation (\ref{tetrad diagonal form}) into the field equations \eqref{field equations}, we get the following modified Friedmann equations
\begin{align}
    3H^2=\kappa\rho-6H^2 f_T+\frac{1}{2}f_{\mathcal{L}_m}\left(\rho+{\mathcal{L}_m}\right)-\frac{f}{2}\, ,
    \label{first Friedmann}
\end{align}
\begin{align}\label{second Friedmann}
        \Dot{H}=\frac{-4\pi G(\rho+p)-Hf_{T{\mathcal{L}_m}}\partial_t{\mathcal{L}_m}
        -\frac{1}{2}f_{\mathcal{L}_m}\left(p+{\mathcal{L}_m}\right)}{1+(f_T-12H^2f_{TT})},
\end{align}
where $H$ is the Hubble parameter with $H=\dot{a}/{a}$. The pressure and energy density are given by $p$ and $\rho$, respectively. 
By using the equations (\ref{tetrad}), (\ref{Torsion}), and (\ref{torsion scalar}), the simplified torsion scalar in terms of the Hubble parameter for the FLRW metric is of the following form,
\begin{align}
    T=-6H^2\,.
    \label{T}
\end{align}

One can rewrite the above modified Friedmann equations in a GR-equivalent form as
\begin{align}
    3 H^2= \kappa\left(\rho+\rho_{DE}\right)\label{Hubble}\, ,
\end{align}
\begin{align}
    2\dot{H}+3H^2=-\kappa \left(p+p_{DE}\right)\, ,
\end{align}
where
\begin{align}
    \rho_{DE}=\frac{3}{\kappa}\left[-2H^2f_T+\frac{1}{6}f_{\mathcal{L}_m}\left(\rho+\mathcal{L}_m\right)-\frac{f}{6}\right]\label{rho_{DE}}\, ,
\end{align}
and
\begin{align}
       p_{DE}=-\rho_{DE}+\frac{1}{\kappa}\Big[2\Dot{H}(f_T-12H^2f_{TT})\nonumber\\+2H\partial_t(\mathcal{L}_m)f_{T{\mathcal{L}_m}}+f_{\mathcal{L}_m}(p+\mathcal{L}_m)\Big]\, .\label{Constrain1}
\end{align}

To determine the continuity equation, we use the Lagrangian ${\mathcal{L}_m}=p$. Choosing instead ${\mathcal{L}_m}=-\rho$, leads to an additional force, which directly arises from the effect of the coupling mechanism. Nevertheless, even with ${\mathcal{L}_m}=p$, the persistence of the energy momentum source term indicates that the gravitational sector remains intrinsically non-conservative \cite{KavyaVenkatesha:2024bpj}. 
The effective equation of state (EoS) parameter can
be defined as
\[
w_{Tot}(z) = \frac{p_{Tot}(z)}{\rho_{Tot}(z)},
\]
where $p_{Tot}(z)= p+p_{DE}$ and $\rho_{Tot}(z)=\rho+\rho_{DE} $.

The simplest realization of inflation is provided by GR
minimally coupled to a homogeneous scalar field $\phi$, the inflaton,
governed by the action
\begin{align}
S=\int d^4x\,\sqrt{-g}\left(-\frac{R}{2\kappa}
+\mathcal{L}_m^{[\phi]}\right),
\label{eq:action}
\end{align}
where $\kappa=1/M_{\rm Pl}^2$, $M_{\rm Pl}$ denotes the
reduced Planck mass, $g$ is the determinant of the metric, and
\begin{align}
\mathcal{L}_m^{[\phi]}=\frac{1}{2}g^{\mu\nu}\partial_\mu\phi\,
\partial_\nu\phi-V(\phi),
\label{eq:Lm}
\end{align}
with $V(\phi)$ the inflaton potential. 

The energy-momentum tensor is
defined as
\begin{align}
T_{\mu\nu}\equiv\frac{2}{\sqrt{-g}}
\frac{\delta\left(\sqrt{-g}\,\mathcal{L}_m\right)}{\delta g^{\mu\nu}},
\label{eq:Tdef}
\end{align}
which, upon insertion of Eq.~\eqref{eq:Lm}, gives
\begin{align}
T_{\mu\nu}^{[\phi]}=\partial_\mu\phi\,\partial_\nu\phi
-g_{\mu\nu}\left[\frac{1}{2}\partial^\sigma\phi\,\partial_\sigma\phi
-V(\phi)\right].
\label{eq:Tmunu}
\end{align}

Varying Eq.~\eqref{eq:action} with respect to the metric and using
Eq.~\eqref{eq:Tdef} then yields the Einstein field equations
\begin{align}
G_{\mu\nu}\equiv R_{\mu\nu}-\frac{1}{2}R\,g_{\mu\nu}=\kappa\,T_{\mu\nu}.
\label{eq:EFE}
\end{align}
Considering the spatially flat FLRW geometry and comparing Eq.~\eqref{eq:Tmunu} with the perfect-fluid form
$T_{\mu\nu}=(\rho+p)u_\mu u_\nu-p\,g_{\mu\nu}$, where $u_\mu u^\mu=+1$,
gives rise to the energy density and pressure of the inflaton as
\begin{align}
\rho^{[\phi]}=\frac{1}{2}\dot\phi^2+V(\phi),
\qquad
p^{[\phi]}=\frac{1}{2}\dot\phi^2-V(\phi).
\label{eq:rhop}
\end{align}

The independent components of Eq.~\eqref{eq:EFE} then reduce to the
Friedmann equations
\begin{align}
H^2&=\frac{\kappa}{3}\left(\frac{\dot\phi^2}{2}+V\right),
\label{eq:F1}\\
\dot H&=-\frac{\kappa}{2}\dot\phi^2,
\label{eq:Hdot}
\end{align}
while variation of Eq.~\eqref{eq:action} with respect to $\phi$ gives the
Klein--Gordon equation
\begin{align}\label{KG_GR}
\ddot\phi+3H\dot\phi+V'=0,
\end{align}
where ``dot" represents derivative with respect to cosmic time and ``prime" represents derivative with respect to $\phi$, the inflaton field.

To describe how the cosmological background evolves, one can introduce a series of parameters called the Hubble flow functions, written as $\epsilon_n$
\cite{Hoffman:2000ue,Schwarz:2001vv}
\begin{align}
\epsilon_{n+1}\equiv\frac{d\ln|\epsilon_n|}{dN},
\qquad n\geq0,
\label{eq:HFF}
\end{align}
with $\epsilon_0\equiv H_{\rm i}/H$ and $N=log(\frac{a}{a_i})$ is the number of e-folds. The first two members read
\begin{align}
\epsilon_1=-\frac{\dot H}{H^{2}},
\qquad
\epsilon_2=\frac{\dot\epsilon_1}{H\epsilon_1}.
\label{eq:eps12}
\end{align}

Sustained inflation requires $|\epsilon_n|\ll1$ \cite{Liddle:1994dx} for
all $n$, and the accelerated phase terminates when $\epsilon_1=1$.
\newline
The slow-roll scenario occurs in the regime  
\begin{align}\label{slorroll condn}
\dot\phi^{2}\ll V \quad \mbox{and}
\quad |\ddot\phi|\ll3H|\dot\phi|,    
\end{align}
 which approximates the first FLRW equation \eqref{eq:F1} and the scalar field equation \eqref{eq:Hdot} as,
\begin{align}
H^{2}\simeq\frac{\kappa}{3}V,
\qquad
3H\dot\phi\simeq-V',
\label{eq:SRbg}
\end{align}
so that the leading Hubble-flow functions admit the standard potential representation \cite{Liddle:1994dx,Martin:2013tda}
\begin{align}
\epsilon_1\simeq\frac{1}{2\kappa}\left(\frac{V'}{V}\right)^{2},
\qquad
\epsilon_2\simeq\frac{2}{\kappa}\left[\left(\frac{V'}{V}\right)^{2}
-\frac{V''}{V}\right].
\label{eq:potSR}
\end{align}

To first order in these quantities, the scalar spectral index and the
tensor-to-scalar ratio evaluated at horizon crossing are
\cite{Stewart:1993bc,Martin:2013tda}
\begin{align}
n_{\rm s}=1-2\epsilon_1-\epsilon_2,
\label{eq:ns}
\end{align}
\begin{align}\label{eq:r}
    r=16\epsilon_1.
\end{align}

The number of e-folds \cite{Baumann:2009ds} remaining before inflation ends, i.e., between horizon crossing at
$\phi=\phi_*$ and the end of inflation at
$\phi=\phi_{\rm end}$ follows as
\begin{align}\label{eq:Nefolds}
N \equiv \ln\frac{a_{\rm end}}{a}
= \int_{t}^{t_{\rm end}} H\,dt
= \int_{\phi_{\star}}^{\phi_{\rm end}} \frac{H}{\dot\phi}\,d\phi.
\end{align}

By using the slow-roll relations from equation \eqref{eq:SRbg}, we get
\begin{align}
  N  \approx \kappa\int_{\phi_{\rm end}}^{\phi_{\star}} \frac{V}{V'}\,d\phi .
\end{align}

Observationally viable models typically require $N_*\simeq50$--$60$
\cite{Liddle:2003as,Planck:2018X}.

\section{SLOW-ROLL INFLATION IN $f(T, \mathcal{L}_m)$
GRAVITY WITH SCALAR FIELD}\label{sec:slowroll}
\begin{figure}
    \centering
    \includegraphics[width=\linewidth]{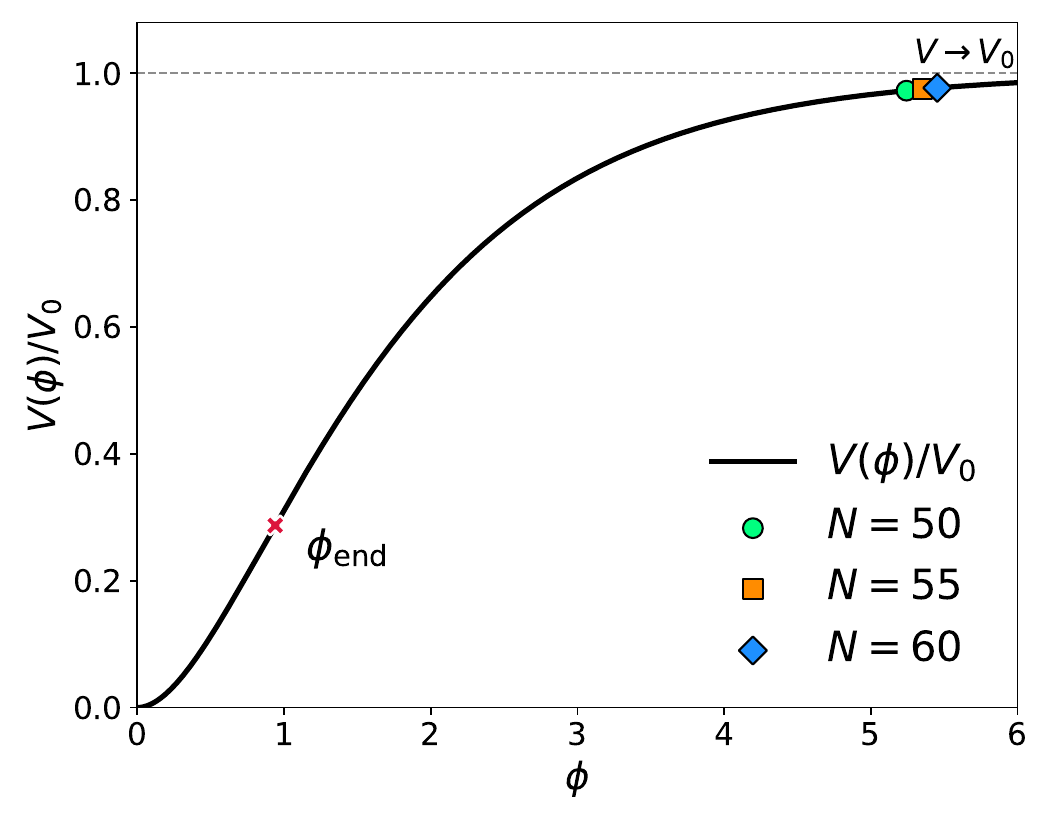}
    \caption{The Starobinsky inflationary potential $V(\phi)/V_0$ as a function of
the inflaton field $\phi$. The red marker indicates the end of
inflation at $\phi_{\rm end}$, while the markers at $N=50$, $55$, and
$60$ denote the corresponding horizon-exit field values in the
$\alpha=\beta=0$ Starobinsky benchmark. The horizon-exit points lie
on the asymptotic plateau, $V(\phi)\rightarrow V_0$, with larger
numbers of $e$-folds corresponding to larger values of $\phi_\star$.}
    \label{fig:potential}
\end{figure}

\begin{figure}
    \centering
    \includegraphics[width=\linewidth]{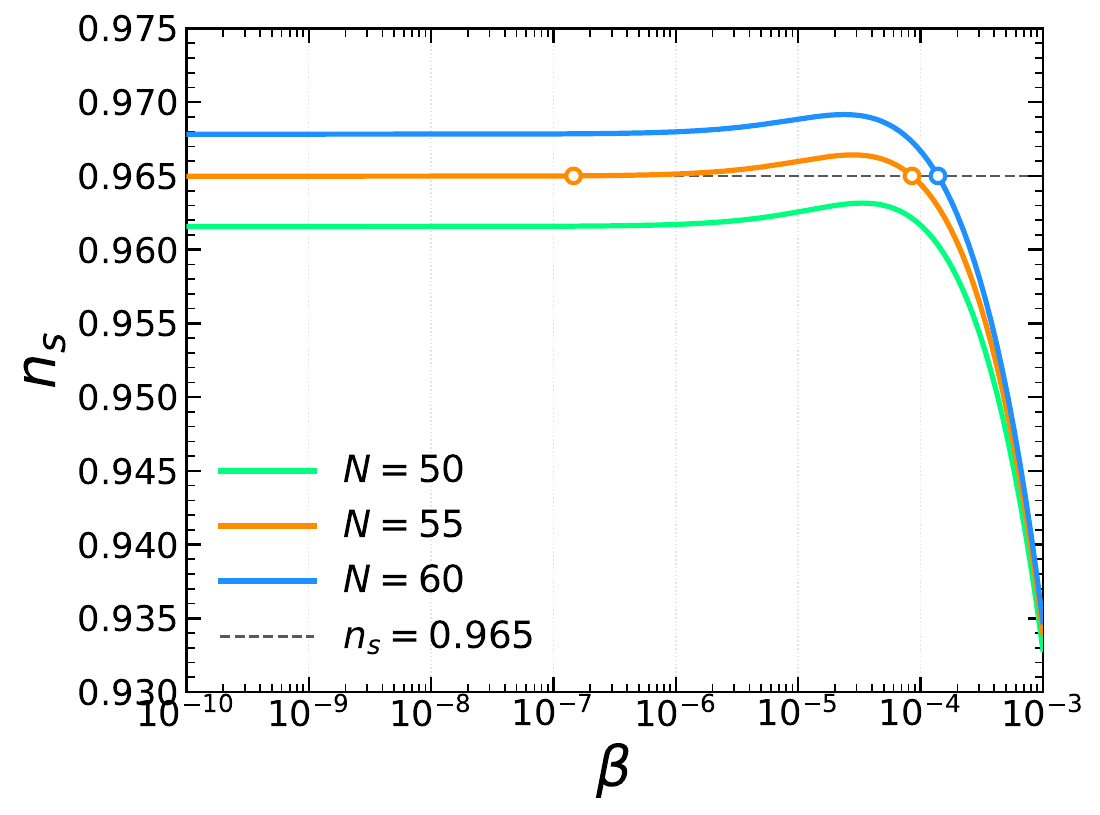}
    \caption{Scalar spectral index $n_s$ as a function of the matter-geometry coupling parameter $\beta$ with $\alpha=0$, shown for $N=50$, $55$, and $60$ e-folds. The horizontal dashed line denotes the reference value $n_s=0.965$, while the open markers indicate the corresponding intersection points of the model predictions with this reference value. For $N=55$, two intersections occur at $\beta\simeq1.46\times10^{-7}$ and $\beta\simeq8.52\times10^{-5}$, whereas for $N=60$ a single intersection occurs at $\beta\simeq1.39\times10^{-4}$. No intersection is found for $N=50$ within the displayed $\beta$ range.}
    \label{fig:ns_beta}
\end{figure}
\begin{figure}
    \centering
    \includegraphics[width=\linewidth]{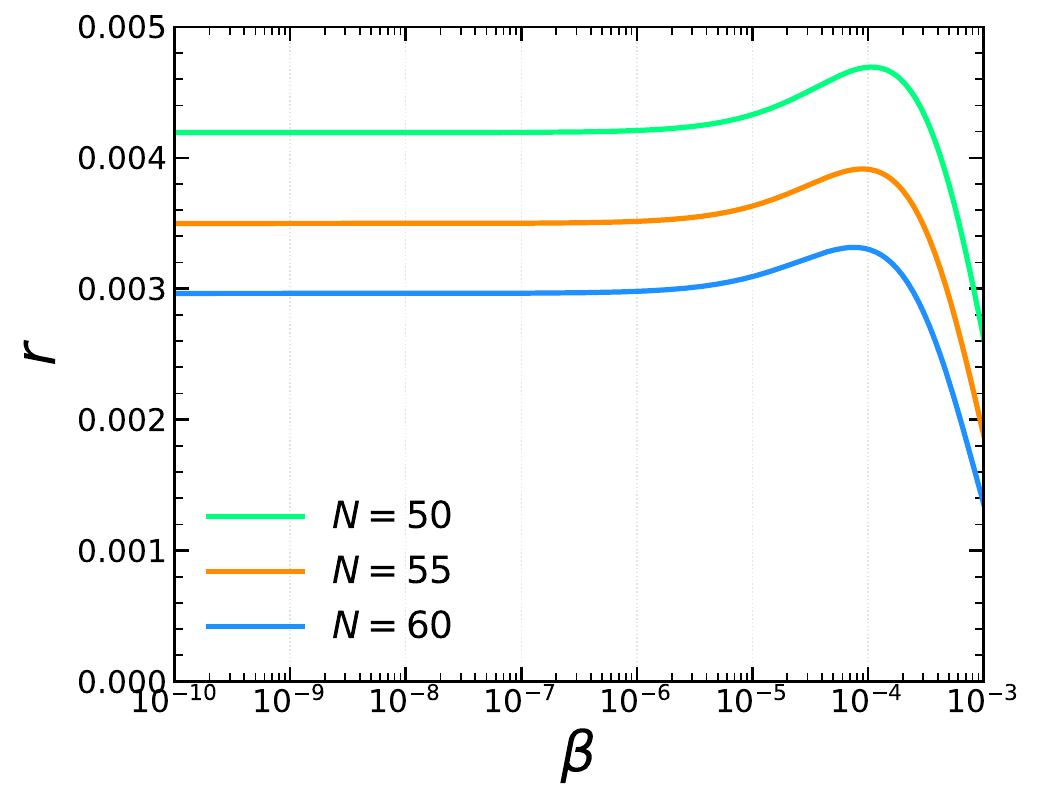}
    \caption{Tensor-to-scalar ratio $r$ as a function of the model parameter $\beta$ for the Starobinsky $f(T, L_m)$ inflationary scenario, shown for $N=50$, $55$, and $60$ e-folds with $\alpha=0$. The figure illustrates the dependence of the tensor-to-scalar ratio on $\beta$ and the shift in the predicted tensor amplitude with the number of e-folds.}
    \label{fig:r_beta}
\end{figure}
\begin{figure*}
    \centering
    \includegraphics[width=0.69\linewidth]{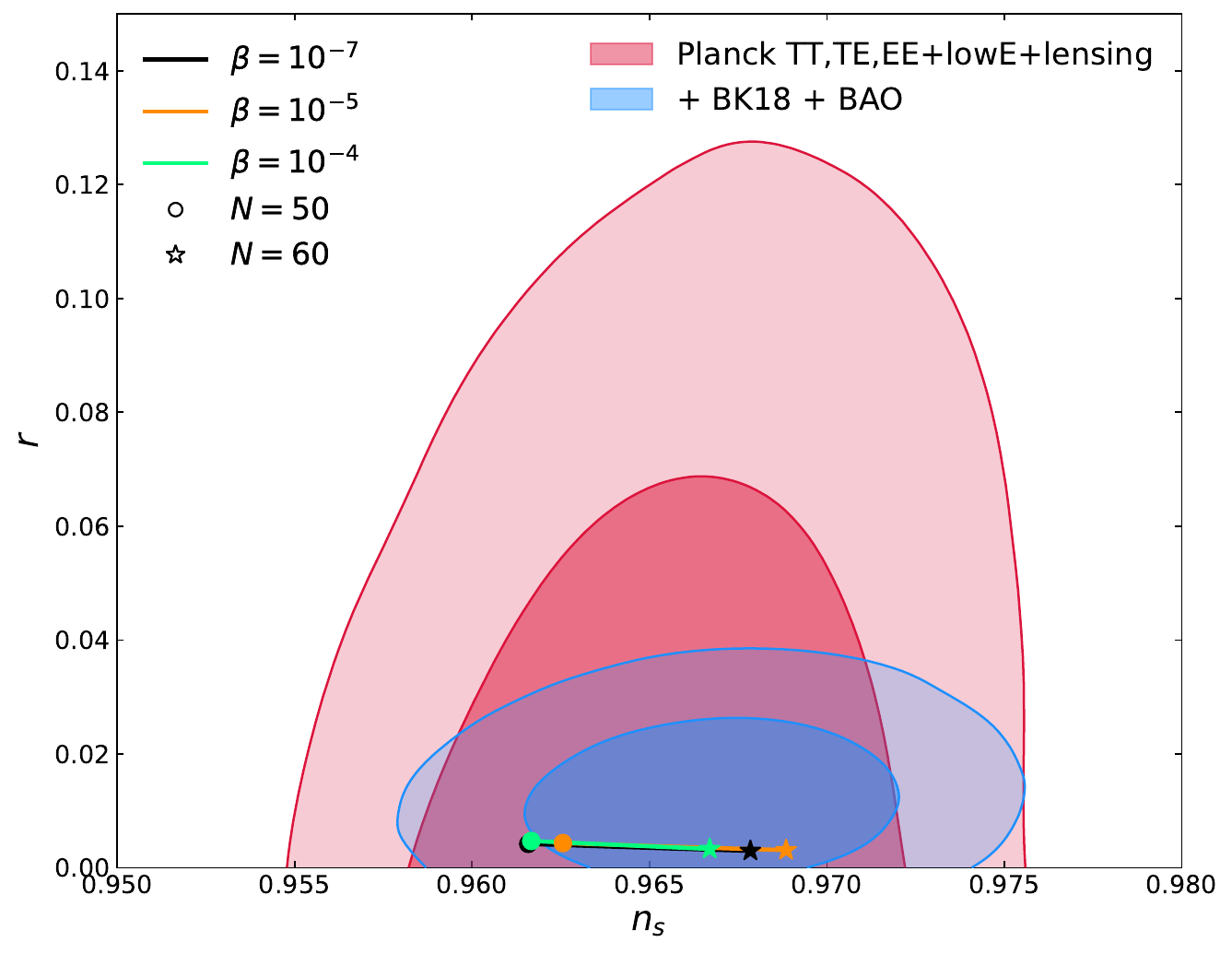}
    \caption{The scalar spectral index $n_s$--tensor-to-scalar ratio $r$ parameter space for the Starobinsky potential  $f(T,L_m)$ inflationary model. The Planck TT,TE,EE+lowE+lensing and Planck TT,TE,EE+lowE+lensing+BK18+BAO observational constraints are shown by the corresponding $1\sigma$ and $2\sigma$ confidence contours. The theoretical trajectories show the evolution of the model predictions as the number of $e$-folds varies from $N=50$ to $N=60$, for three representative values of the coupling parameter, $\beta=10^{-7}$, $10^{-5}$, and $10^{-4}$.}
    \label{fig:ns_r}
\end{figure*}
\begin{figure}
    \centering
    \includegraphics[width=\linewidth]{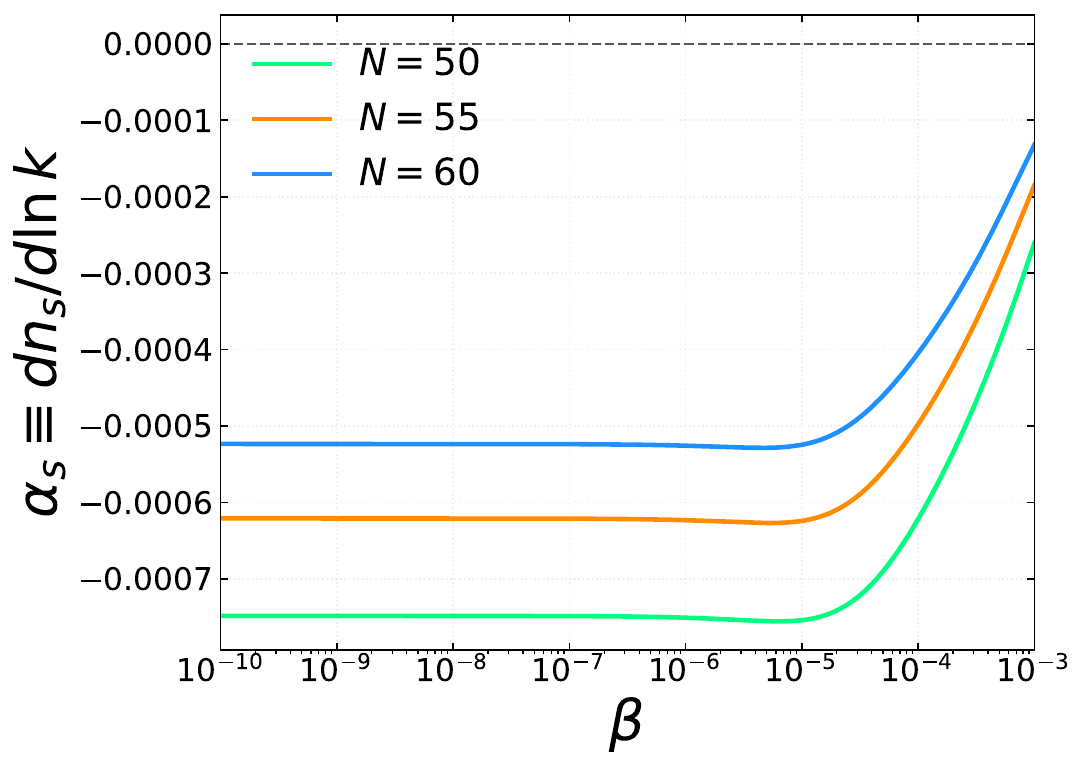}
    \caption{Running of the scalar spectral index, $\alpha_s \equiv dn_s/d\ln k$, as a function of $\beta$ for $N=50$, $55$, and $60$. The curves show the characteristic dependence of the running on $\beta$, with $\alpha_s$ remaining negative throughout the parameter range considered.}
    \label{fig:alphas_beta}
\end{figure}
\begin{table*}
\caption{%
GR Starobinsky benchmark obtained in the $\beta\rightarrow0$ limit
with $\alpha=0$ and $\lambda=\sqrt{2/3\kappa^2}$.
}
\label{tab:GR_benchmark}
\begin{ruledtabular}
\begin{tabular}{c c c c c c c c}
$N$ &
$\phi_{\rm end}$ &
$\phi_*$ &
$\epsilon_1$ &
$\epsilon_2$ &
$n_s$ &
$r$
\\
\hline
50 &
0.94017755 &
5.24351055 &
$2.620192\times10^{-4}$ &
$3.790633\times10^{-2}$ &
0.96156963 &
$4.192307\times10^{-3}$
\\
55 &
0.94017755 &
5.35286852 &
$2.186437\times10^{-4}$ &
$3.458549\times10^{-2}$ &
0.96497722 &
$3.498299\times10^{-3}$
\\
60 &
0.94017755 &
5.45315156 &
$1.852432\times10^{-4}$ &
$3.180238\times10^{-2}$ &
0.96782714 &
$2.963891\times10^{-3}$
\end{tabular}
\end{ruledtabular}
\end{table*}
\begin{table*}
\caption{%
Values of the matter--geometry coupling $\beta$ for which the
scalar spectral index reaches the reference value $n_s=0.965$. For $N=55$, two distinct
solutions occur within the scanned interval $10^{-10}\leq\beta\leq10^{-3}$.
No solution is found for $N=50$ in this interval.%
}
\label{tab:ns_constraint}
\begin{ruledtabular}
\begin{tabular}{c c c c c c}
$N$ &
Solution &
$\beta$ &
$\epsilon_1$ &
$\epsilon_2$ &
$r$
\\
\hline
50 &
-- &
-- &
-- &
-- &
-- 
\\[2pt]
55 &
1 &
$1.459571\times10^{-7}$ &
$2.187910\times10^{-4}$ &
$3.456242\times10^{-2}$ &
$3.500656\times10^{-3}$
\\
55 &
2 &
$8.528572\times10^{-5}$ &
$2.446452\times10^{-4}$ &
$3.451071\times10^{-2}$ &
$3.914324\times10^{-3}$
\\
60 &
1 &
$1.389661\times10^{-4}$ &
$2.025442\times10^{-4}$ &
$3.459491\times10^{-2}$ &
$3.240707\times10^{-3}$
\end{tabular}
\end{ruledtabular}
\end{table*}
\begin{table*}
\caption{%
The maximum value of the scalar spectral index obtained by varying
$\beta$ for each $N$.
The corresponding coupling, horizon-crossing field value,
slow-roll parameters, and tensor-to-scalar ratio are also shown.%
}
\label{tab:ns_maximum}
\begin{ruledtabular}
\begin{tabular}{c c c c c c c}
$N$ &
$\beta_{\rm max}$ &
$\phi_*$ &
$\epsilon_1$ &
$\epsilon_2$ &
$n_{s,\rm max}$ &
$r$
\\
\hline
50 &
$3.346909\times10^{-5}$ &
5.38069715 &
$2.829082\times10^{-4}$ &
$3.627700\times10^{-2}$ &
0.96315718 &
$4.526530\times10^{-3}$
\\
55 &
$2.793243\times10^{-5}$ &
5.49033161 &
$2.360453\times10^{-4}$ &
$3.310090\times10^{-2}$ &
0.96642701 &
$3.776725\times10^{-3}$
\\
60 &
$2.366812\times10^{-5}$ &
5.59084689 &
$1.999655\times10^{-4}$ &
$3.043880\times10^{-2}$ &
0.96916127 &
$3.199447\times10^{-3}$
\end{tabular}
\end{ruledtabular}
\end{table*}
\begin{table*}
\caption{%
Running of the scalar spectral index for representative
values of $\beta$ and $N$. The results are numerically stable
with respect to the finite-difference step used to evaluate
$dn_s/dN$.%
}
\label{tab:running}
\begin{ruledtabular}
\begin{tabular}{c c c c}
$N$ &
$\beta$ &
$\alpha_s$
\\
\hline
50 & $10^{-10}$ & $-7.482418\times10^{-4}$ \\
50 & $10^{-7}$  & $-7.485087\times10^{-4}$ \\
50 & $10^{-5}$  & $-7.539872\times10^{-4}$ \\
50 & $10^{-4}$  & $-6.218748\times10^{-4}$ \\
50 & $10^{-3}$  & $-2.607651\times10^{-4}$ \\
\hline
55 & $10^{-10}$ & $-6.207641\times10^{-4}$ \\
55 & $10^{-7}$  & $-6.210300\times10^{-4}$ \\
55 & $10^{-5}$  & $-6.240833\times10^{-4}$ \\
55 & $10^{-4}$  & $-4.977772\times10^{-4}$ \\
55 & $10^{-3}$  & $-1.849102\times10^{-4}$ \\
\hline
60 & $10^{-10}$ & $-5.233693\times10^{-4}$ \\
60 & $10^{-7}$  & $-5.236343\times10^{-4}$ \\
60 & $10^{-5}$  & $-5.244485\times10^{-4}$ \\
60 & $10^{-4}$  & $-4.048043\times10^{-4}$ \\
60 & $10^{-3}$  & $-1.321353\times10^{-4}$
\end{tabular}
\end{ruledtabular}
\end{table*}
We will now focus on a specific model defined by the equation 
\begin{align}\label{model}
    f(T,\mathcal{L}_m)=\alpha T^2 +\beta \mathcal{L}_m
\end{align}
where $\alpha$ and $\beta$ are constants. The quadratic torsion term modifies
the gravitational sector at high energies, while $\beta$ controls the strength
of the torsion matter coupling. Setting $\alpha = \beta = 0$ recovers
Teleparallel gravity with a minimally coupled scalar field.

Substituting Eq.~\eqref{model} in Eq.~\eqref{first Friedmann} and using Eq.~\eqref{torsion scalar} gives
\begin{align}\label{quartic}
    54\alpha H^4-3H^2+A\rho^{[\phi]}=0,
    \qquad A \equiv \kappa + \frac{\beta}{2} ,
\end{align}
which is quadratic in $H^2$ and admits the two branches
\begin{align}\label{quartic2}
    H^2=\frac{1\pm \sqrt{1-24\alpha A \rho^{[\phi]}}}{36\alpha}.
\end{align}

The upper branch diverges as $\alpha \to 0$ and is therefore unphysical. We retain the lower branch, which is continuously connected to GR.
Expanding it to the second order in the small quantity
$24\alpha A \rho^{[\phi]} \ll 1$ yields
\begin{align}\label{H2approx}
H^{2} \simeq \frac{A\rho^{[\phi]}}{3}\left(1 + 6\alpha A \rho^{[\phi]}\right).
\end{align}

The leading term reproduces the standard Friedmann equation, and the correction
is first order in $\alpha$.

The second Friedmann equation \eqref{second Friedmann} takes the form
\begin{align}\label{2nd Friedmann}
    \dot{H}=\frac{(-4\pi G-\beta/2)\dot{\phi}^2 +\beta V}{1-36\alpha H^2}.
\end{align}

Differentiating Eq.~\eqref{quartic} with respect to cosmic time and combining the
result with Eq.~\eqref{2nd Friedmann}, we obtain the modified Klein-Gordon equation
\begin{align}\label{KG}
    A\dot{\phi}\ddot{\phi}+3HB\dot{\phi}^2+ A\dot{\phi}V'=6H\beta V,
\end{align}
where from now on we take $B=(\kappa +\beta)$. In the limit $\beta\rightarrow0$, both $A$ and $B$ reduce to $\kappa$, and Eq.~\eqref{KG} recovers the standard Klein-Gordon equation of GR (Eq.~\eqref{KG_GR}). For $\beta \neq 0$, the coupling modifies the scalar field dynamics in two ways. First, the Hubble friction is rescaled by $B/A$, so for $\beta> 0$ the damping of the field is slightly enhanced relative to GR. Second, the term $6H\beta V$ acts as a GR counterpart, representing energy exchange between the scalar field and the geometry, so the scalar field is no longer separately conserved. Since this source is proportional to both $H$ and $V$, it is most relevant during slow roll, where the potential energy dominates.

Imposing the slow roll conditions \eqref{slorroll condn}, 
Eqs.~\eqref{H2approx} and ~\eqref{KG} reduce to
\begin{align}\label{H2sr}
H^{2} \simeq \frac{AV}{3}\left(1 + 6\alpha A V\right),
\end{align}
\begin{align}\label{phidot}
\dot{\phi} = \frac{-AV' - \sqrt{(AV')^{2} + 72\beta B H^{2} V}}{6HB},
\end{align}
where the sign of the square root is again fixed by requiring the correct
general relativistic limit.

Using Eq.~\eqref{2nd Friedmann} for $\dot H$ and then eliminating $\dot\phi$ and $H$ by
means of Eq.~\eqref{phidot} and \eqref{H2sr}, the first slow roll parameter can be
expressed entirely in terms of the potential,
\begin{align}\label{eps1}
\epsilon_1=\frac{V' \left(A V'+ \sqrt{A \left( 144 \alpha \beta A B V^3 + A V'^2 + 24 \beta B V^2 \right)}  \right)}{4 A B V^2 \left( 6 \alpha A V + 1 \right)^2 \left( 1-12 \alpha A V- 72 \alpha^2 A^2 V^2  \right)}.
\end{align}
Proceeding in the same way for $\epsilon_2$ gives
\begin{align}\label{epsilon2}
\epsilon_{2} &=
\frac{\sqrt{A^{2}V'^{2}+24\beta ABV^{2}(6\alpha AV+1)}+AV'}
     {2ABV(6\alpha AV+1)} \notag\\[4pt]
&\times\biggl[\,
\frac{V'\bigl(9\alpha AV(4\alpha AV(54\alpha AV+13)+1)-1\bigr)}
     {V(6\alpha AV+1)\bigl(12\alpha AV(6\alpha AV+1)-1\bigr)} \notag\\[4pt]
&\quad-\frac{V''}{V'}
-\frac{AV''}{\sqrt{A\bigl(24\beta BV^{2}(6\alpha AV+1)+AV'^{2}\bigr)}} \notag\\[4pt]
&+\frac{AV'^{2}(9\alpha AV+1)}
{V(6\alpha AV+1)\sqrt{A\bigl(24\beta BV^{2}(6\alpha AV+1)+AV'^{2}\bigr)}}
\,\biggr].
\end{align}

Both expressions reduce to their standard forms in the limit
$\alpha, \beta \to 0$.

Substituting Eq.~\eqref{H2sr} and Eq.~\eqref{phidot} into Eq.~\eqref{eq:Nefolds}, the number of e-folds becomes
\begin{align}\label{efoldinV}
    N=\int_{\phi_{end}}^{\phi_{\star}} \frac{2ABV(1+6\alpha AV)}{AV'+\sqrt{(AV')^2+24\beta ABV^2(1+6\alpha AV)}} d \phi.
\end{align}

As a benchmark, we adopt the Starobinsky-type potential \cite{Starobinsky:1980te}
\begin{align}\label{potentail}
V(\phi)= V_0\left(1 - e^{-\lambda\phi}\right)^2,
\end{align}
which reduces to the canonical potential obtained from the conformal transformation of $R^2$ gravity to the Einstein frame when $\lambda = \sqrt{2/3\kappa^2}  $ \cite{Starobinsky:1980te}. The defining feature of this potential is its plateau at large field values, $V(\phi)\to V_0 \quad \mbox{as}\quad \phi\to\infty$. Because this energy potential flattens out at high energies, it drives cosmic inflation that perfectly matches observational data \cite{Planck:2018X}. For illustration, Fig.~\ref{fig:potential} shows the Starobinsky
potential together with the end-of-inflation field value and the
horizon-exit field values corresponding to $N=50$, $55$, and $60$ in
the GR benchmark limit.

For Starobinsky Potential Eq.~\eqref{potentail}, the first slow-roll parameter transforms as
\begin{align}\label{epsilon1inVform}
\epsilon_1 = \frac{e^{-\lambda \phi} \lambda \left(A \lambda e^{-\lambda \phi} +\mathcal{P}\right)}{A B \mathcal{U} \mathcal{Q}^2 \left(1-12 AV_0 \alpha \mathcal{U}\mathcal{Q}\right)}.
\end{align}

Similarly, the second slow-roll parameter reads as
\begin{align}\label{epsilon2inVform}
\epsilon_{2} =&{}
\frac{A\lambda e^{-\lambda\phi}+\mathcal{P}}{AB\,\mathcal{U}\mathcal{Q}} 
\times\biggl(\,
\frac{2A\lambda^{2}e^{-2\lambda\phi}\left(9\alpha AV_{0}\mathcal{U}+1\right)}
{\mathcal{Q}\,\mathcal{P}} \notag\\[4pt]
& -\frac{AV''}{2\mathcal{P}}
-\frac{e^{\lambda\phi}V''}{2\lambda V_{0}}+ \frac{2\lambda e^{-\lambda\phi}\mathcal{R}}
{\mathcal{Q}\left(12\mathcal{R}\alpha AV_{0}\mathcal{U}\mathcal{Q}-1\right)}
\,\biggr).
\end{align}
By Eq.~\eqref{efoldinV}, the modified e-fold number for the given potential Eq.~\eqref{potentail},
\begin{align}
N =\bigint_{\phi_{end}}^{\phi_{\star}}  \frac{A B\left(1-e^{-\lambda \phi}\right)
\mathcal{Q}}
{\mathcal{P}
+A \lambda e^{-\lambda \phi}}d \phi,
\end{align}
and from substitution of Eq.~\eqref{epsilon1inVform} in Eq.~\eqref{eq:r},
\begin{align}
     r= \frac{16e^{-\lambda \phi} \lambda \left(A \lambda e^{-\lambda \phi} +\mathcal{P}\right)}{A B \mathcal{U} \mathcal{Q}^2 \left(1-12 AV_0 \alpha \mathcal{U}\mathcal{Q}\right)}.
 \end{align}
 
By substitution of slow-roll parameters $\epsilon_1$ and $\epsilon_2$, the spectral index in the form of Starobinsky Potential reads as, 
\begin{align}\label{spectralindexinVform}
n_{s} ={}& 1
-\frac{2e^{-\lambda\phi}\lambda\left(A\lambda e^{-\lambda\phi}+\mathcal{P}\right)}
{AB\,\mathcal{U}\mathcal{Q}^{2}\left(1-12AV_{0}\alpha\mathcal{U}\mathcal{Q}\right)} -\frac{A\lambda e^{-\lambda\phi}+\mathcal{P}}{AB\,\mathcal{U}\mathcal{Q}} \notag\\[4pt]
&\times\biggl(\,
-\frac{AV''}{2\mathcal{P}}
-\frac{e^{\lambda\phi}V''}{2\lambda V_{0}}+\frac{2A\lambda^{2}e^{-2\lambda\phi}\left(9\alpha AV_{0}\mathcal{U}+1\right)}
{\mathcal{Q}\,\mathcal{P}} \notag\\[4pt]
&\qquad+\frac{2\lambda e^{-\lambda\phi}\mathcal{R}}
{\mathcal{Q}\left(12\alpha AV_{0}\mathcal{U}\mathcal{Q}-1\right)}
\,\biggr),
\end{align}

where, we have used the following substitution in above expressions,
\begin{align*}
   \mathcal{U}& =\left(1-e^{-\lambda \phi}\right)^2, \qquad \mathcal{Q}=\left(1+6 A V_0 \alpha \mathcal{U}\right)\\ 
\mathcal{P}&=\sqrt{6 A B \beta \mathcal{U}+ 36 A^2 B \alpha \beta V_0\mathcal{U}^2+ A^2 \lambda^2 e^{-2\lambda \phi}}\\
\mathcal{R}&= \left(9 \alpha A V_0 \mathcal{U} \left(4 \alpha A V_0 \mathcal{U} \left(54 \alpha A V_0 \mathcal{U}+13\right)+1\right)-1\right).
\end{align*}

\section{Interpretations}\label{sec:interpretaion}
The phenomenological analysis presented in this manuscript focuses on the
matter--geometry coupling parameter $\beta$ and is performed in the
$\alpha=0$ sector of the model
$\alpha T^2+\beta\mathcal{L}_m$.
It is important to emphasize that setting $\alpha=0$ does not reduce the
model to GR when $\beta\neq0$. The matter--torsion coupling
associated with $\beta$ remains present in both the background dynamics and
the scalar-field evolution. The $\alpha=0$ choice therefore defines a
nontrivial one-parameter sector of the full $f(T,\mathcal{L}_m)$ theory. The dependence of the inflationary observables
on non-zero $\alpha$ has been separately discussed in Subsection \ref{section:alpha} in detail. The results show that the $T^2$ contribution
produces only small changes in the inflationary observables within the
low-$\alpha$ regime considered here, while its effect becomes
increasingly visible for larger values of $\alpha$, particularly in the
tensor-to-scalar ratio. We therefore adopt $\alpha=0$ in the main
phenomenological analysis as a deliberate one-parameter sector of the
full theory, allowing the effects of the matter--torsion coupling
$\beta$ to be examined without obscuring them by an additional
parameter dependence.

Table \ref{tab:GR_benchmark} establishes the GR benchmark, which is obtained when $\beta\rightarrow0$ with $\alpha=0$ and $\lambda=\sqrt{2/3\kappa^2}$, which serves as the reference for the effect of the couplings. The scalar field value at the end of inflation, $\phi_{end}\approx 0.9402$, is independent of $N$, since it is fixed solely by $\epsilon_1=1$. But the horizon crossing value $\phi_{\star}$ changes from $5.2435- 5.4532$ for $N=50- 60$, with $\epsilon_1=(1.85 - 2.62) \times 10^{-4}$ and $\epsilon_2=(3.18- 3.79) \times 10^{-2}$. The hierarchy $\epsilon_1 \ll \epsilon_2$ is characteristic of plateau inflation and gives through  $n_s=1-2\epsilon_1-\epsilon_2$ and $r=16\epsilon_1$. These values lie $(1.2-1.6)\times10^{-3}$ above the
leading-order estimate $n_s\simeq1-2/N$ and approximately
$11\%-13\%$ below the leading-order prediction $r\simeq12/N^2$,
reflecting the finite-$N$ corrections captured by the numerical
slow-roll treatment \cite{Planck:2018X}.
The $N=55$ value reproduces the Planck 2018 TT,TE,EE+lowE+lensing central value $n_s=0.9649\pm 0.0042$ \cite{Planck:2018VI} to within $10^{-4}$.

In Fig. \ref{fig:ns_beta} three different regimes are shown. For $\beta \lesssim 10^{-7}$, the predictions agree with GR to within $\sim 10^{-5}$, confirming that the model constantly reduces to conventional Starobinsky inflation. At intermediate couplings, $n_s$ progressively increases and reaches a
maximum at $\beta\sim3\times10^{-5}$. For $\beta\gtrsim10^{-4}$ it declines
rapidly, reaching $n_s\approx0.93$ at $\beta=10^{-3}$. The effect of the coupling on the dynamics is considerably more significant for extremely tiny $\beta$ since the potential plateau enhances its contribution to $\epsilon_1$.

Figure \ref{fig:r_beta} shows the corresponding behaviour of $r$. It is flat at the values of Table \ref{tab:GR_benchmark} for $\beta \lesssim 10^{-6}$, rises to a maximum near $\beta \approx 10^{-4}$, and falls steeply thereafter, to $\approx (1.5 -2)\times 10^{-3}$ at $\beta = 10^{-3}$.
At $\beta_{max}$, where $n_s$ peaks, we find $r =(4.527,\, 3.777,3.1999)\times 10^{-3}$ (Table \ref{tab:ns_maximum}). This is about $8\%$ above the $\beta=0$ benchmark. The ratio $r$ keeps growing beyond this point and it peaks to $r=(4.7,4.0,3.3)\times 10^{-3}$ at a larger $\beta$. Since the peak of $r ( \beta \approx10^{-4})$ comes after the peak of $n_s (\beta \approx 3\times 10^{-5})$, each trajectory in Fig. \ref{fig:ns_r} turns back toward lower $n_s$ before reaching its highest $r$.

Table \ref{tab:ns_maximum} shows that the coupling can rise above its GR value at most $\Delta n_s\approx (1.3 -1.6)\times 10^{-3}$, which decreases with $N$. At the maximum, for e.g., at $N=50$, $\epsilon_1$ increases from $2.62 \times 10^{-4}$ to $2.83 \times 10^{-4}$, while $\epsilon_2$ decreases from $3.79 \times 10^{-2}$ to $3.63 \times 10^{-2}$. Since $n_s=1-2\epsilon_1-\epsilon_2$ and $\epsilon_2 \gg \epsilon_1$, the reduction in $\epsilon_2$ dominates and $n_s$ increases. The tensor-to-scalar ratio correspondingly rises from $4.19\times 10^{-3}$ to  $4.53\times 10^{-3}$. Similar changes can also be seen for $N=55$ and $N=60$.

In the GR limit, $n_s$ is fixed once $N$ is specified. The matter-geometry coupling introduces an additional degree of freedom that allows $n_s$ to be shifted away from its GR value. Table \ref{tab:ns_constraint} locates the crossings of the reference value $n_s = 0.965$, the Planck TT,TE,EE+lowE+lensing central value \cite{Planck:2018VI}.
For $N = 55$ there are two solutions, $\beta = 1.46\times 10^{-7}$ on the flat branch ($r = 3.501\times 10^{-3}$) and $\beta = 8.53\times 10^{-5}$ on the
descending branch ($r = 3.914\times 10^{-3}$) of Fig.~\ref{fig:r_beta}. For $N = 60$ one solution at $\beta = 1.39\times 10^{-4}$ ($r = 3.241\times 10^{-3}$) and for $N = 50$ no solution is there, since $n_{s,\max} = 0.96316 < 0.965$. The reference value is therefore attainable only for $N \gtrsim 55$.

Fig.~\ref{fig:ns_r} compares the predictions for $\beta$ values at $10^{-7},\, 10^{-5}$ and $10^{-4}$ (where $\alpha=0$) with the 68\% and 95\%
contours from Planck TT,TE,EE+lowE+lensing and from the joint Planck TT,TE,EE+lowE+lensing+BK18+BAO analysis \cite{BICEP:2021xfz}. Both $n_s$ and $r$ are evaluated at the horizon-crossing epoch of the mode that exits $N$ e-folds before the end of inflation.\footnote{The
difference between the $n_s$ and $r$ pivot scales of
Refs.\cite{Planck:2018VI, BICEP:2021xfz} shifts the predictions by
amounts well within the contour widths and is therefore neglected.}

For all three couplings (i.e., for three different $\beta$) the full $N = 50-60$ segment lies inside the
95\% contour of Planck TT,TE,EE+lowE+lensing+BK18+BAO.
The $N = 60$ endpoints lie inside the 68\% contour, while the $N = 50$
endpoints lie at its boundary, so that $N \gtrsim 55$ is preferred at the
$1\sigma$ level, consistent with Table \ref{tab:ns_constraint}. 
For different $\beta$'s, the trajectories are separated by $\Delta n_s \lesssim 3\times 10^{-3}$ and
$\Delta r \lesssim 5\times 10^{-4}$, and their ordering follows
Figs. \ref{fig:ns_beta}- \ref{fig:r_beta}. For $\beta = 10^{-5}$, where two solutions exist, we show the one near the maximum of $n_s$; this trajectory lies at the largest $n_s$. At $\beta = 10^{-4}$,
$n_s$ has fallen back to approximately or slightly below its $\beta \to 0$
value (Table \ref{tab:ns_constraint}: $n_s = 0.965$ is reached at
$\beta = 1.39\times 10^{-4}$ for $N = 60$) while $r$ is near its maximum, so that trajectory is displaced slightly to the left and upward.
For $\beta \lesssim 10^{-5}$ the predictions are indistinguishable from the
Starobinsky limit.
All trajectories lie in the concave region of the plane, far below the
$\phi^{2}$ and $\phi$ predictions, which fall outside the 95\%
contours \cite{Planck:2018VI,BICEP:2021xfz}.
The predicted $r \approx (3.0-4.7)\times 10^{-3}$ is an order of
magnitude below $r_{0.05} < 0.036$ \cite{BICEP:2021xfz}, but at the level of the
$\sigma(r) \sim 0.003$ projected for BICEP Array with delensing \cite{BICEP:2021xfz}
and of the LiteBIRD and CMB-S4 targets \cite{LiteBIRD:2022cnt, CMB-S4:2020lpa}, so this region can be probed by forthcoming $B$-mode measurements.

Table \ref{tab:running} and Fig. \ref{fig:alphas_beta} give
$\alpha_s \equiv {\rm d}n_s/{\rm d}\ln k \simeq -{\rm d}n_s/{\rm d}N$. In the $\beta \to 0$ limit $\alpha_s \text{ as} \quad( -7.48,\, -6.21,\, -5.23 )\times 10^{-4}$ for $N = 50,\, 55,\, 60$,  around $6\%$ below the leading-order Starobinsky estimate $-2/N^{2} \text{ as}\quad (-8.00,\, -6.61,\, -5.56) \times 10^{-4}$ in magnitude, and negative throughout the scanned range. Up to $\beta = 10^{-5}$ the running changes by less than $1\%$ . At $\beta = 10^{-4}$ it is reduced to $(-6.22, -4.98, -4.05) \times 10^{-4}$; at $\beta = 10^{-3}$ it is reduced to ($-2.61,\, -1.85,\, -1.32) \times 10^{-4}$. Over the allowed range $\beta \lesssim 10^{-4}$ the running therefore lies in $-(4.0- 7.5)\times 10^{-4}$. The results are stable against the finite-difference with step size $0.01$.
All values are far inside the Planck constraint $\alpha_s = -0.0045 \pm 0.0067$ \cite{Planck:2018X}, so the running places no additional restriction on $\beta$ at present but is a definite prediction for future measurements.
\subsection{Dependence on $\alpha$}\label{section:alpha}
\begin{figure}
    \centering
    \includegraphics[width=\linewidth]{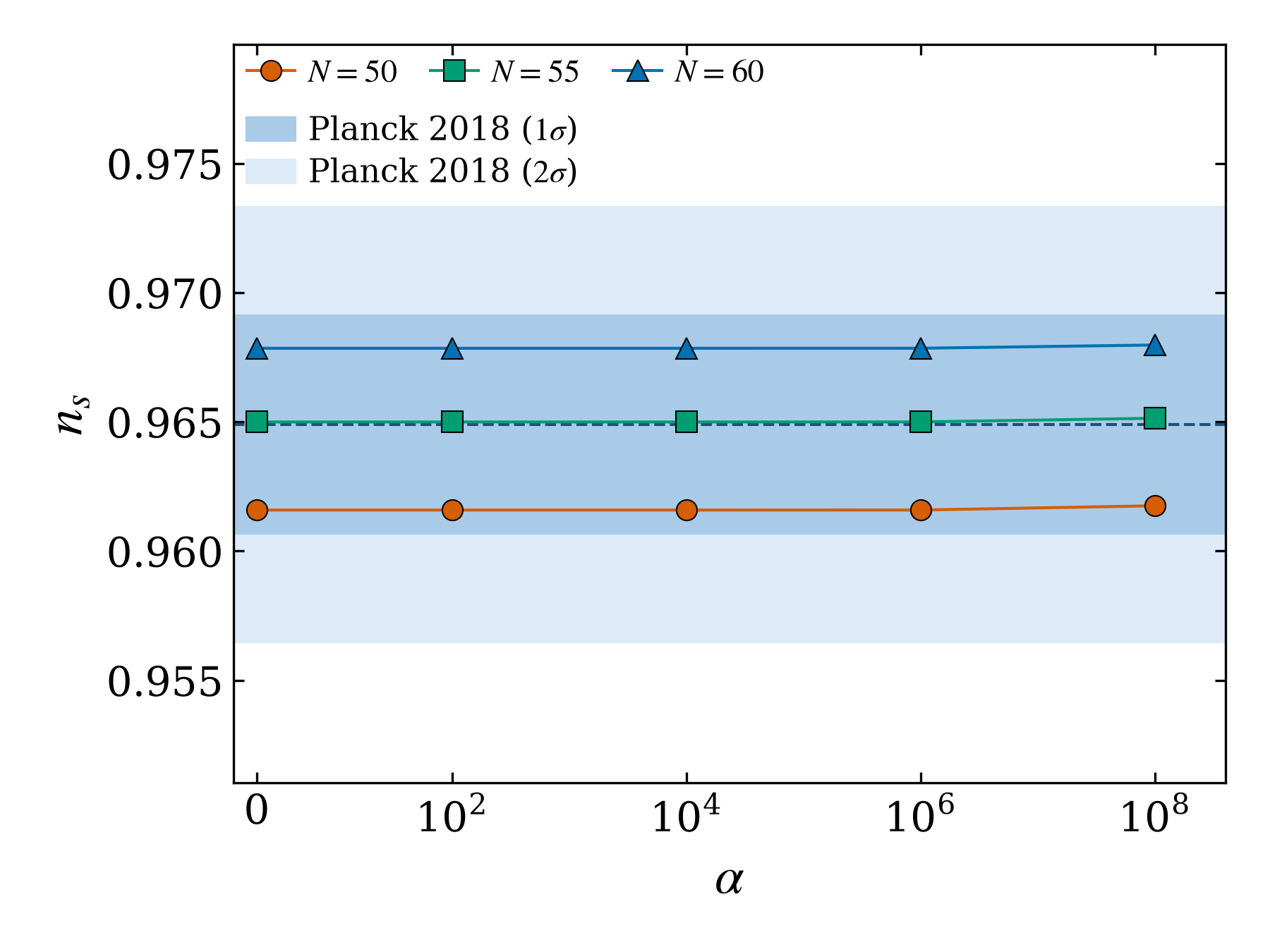}
    \caption{Spectral index $n_s$ versus $\alpha$ for $\beta = 10^{-7}$
and $N = 50, 55, 60$. Shaded regions show the Planck 2018 $1\sigma$
and $2\sigma$ bounds, $n_s = 0.9649 \pm 0.0042$; the dashed line marks
the central value.}
    \label{fig:ns_alppha}
\end{figure}
To assess the impact of the $T^2$ sector independently, we have performed
an explicit numerical scan over $\alpha$ while keeping
$
    \beta=10^{-7}, \
    V_0=10^{-10}, \
    \lambda=\sqrt{2/3\kappa^2},
$
and considering $N=50$, $55$, and $60$ e-folds. The resulting values of
$\phi_\star$, $\epsilon_1$, $\epsilon_2$, $n_s$, and $r$ are reported in
Table \ref{tab:alpha_dependence}.
\par
The $\alpha$ dependence enters the background dynamics through the modified
Friedmann equation \eqref{quartic}
whose solution is given in Eq. \eqref{quartic2}. 
For the branch continuously connected to the standard low-$\alpha$
limit, the approximation employed in the present analysis is valid
when
\begin{equation}\label{lowalpha}
24\alpha A\rho^{[\phi]}\ll1,
\end{equation}
leading to Eq.~\eqref{H2approx}. For the numerical values considered here, we therefore restrict the scan to
$
0\leq\alpha\leq10^8,
$
chosen such that the low-$\alpha$ condition (Eq.\eqref{lowalpha}) remains reasonably controlled for the
inflationary energy scale considered here. Thus, the numerical
importance of $\alpha$ is controlled not by $\alpha$ alone, but by the
dimensionless combination $\alpha A\rho^{[\phi]}$.
These relations also explain why comparatively large numerical values of
$\alpha$ can still produce small corrections for the inflationary energy
scale considered here.

Table~\ref{tab:alpha_dependence} and Fig.~\ref{fig:ns_alppha} shows the effect of $\alpha$ for $0 \le \alpha \le 10^{8}$, the range in which the low $\alpha$ approximation to the background equations is valid. The $\alpha=0$ row differs from Table~\ref{tab:GR_benchmark} by $\Delta n_s =(1.4-1.7)\times 10^{-5}$, consistent with the flat branch of Fig.~\ref{fig:ns_beta}. For $\alpha \le 10^{4}$ the observables are unchanged at the quoted precision
($\Delta n_s \le 2\times 10^{-8}$, $\Delta r/r \approx 10^{-5}$). At
$\alpha = 10^{6}$ the shifts are $\Delta n_s = (1.2 -1.6)\times 10^{-6}$ and $\Delta r/r = +0.1\%$. 

Only at $\alpha = 10^{8}$ is the coupling noticeable for corresponding $N=50- 60$, $\phi_{*}$ decreases by around $1.2\%$, $\epsilon_1$ increases by $12.6\% -12.9\%$ while $\epsilon_2$ decreases by $0.55\% -0.61\%$, so that $r$ rises by $12.6\%-12.9\%$ to ($4.724\,-\, 3.347 )\times 10^{-3}$ , whereas $n_s$ increases by only ($1.66-\, 1.26) \times 10^{-4} \%$ (the decrease of $\epsilon_2$ outweighs the increase of $2\epsilon_1$ in $n_s = 1 - 2\epsilon_1 - \epsilon_2$). 
\par
The $T^{2}$ term therefore acts almost exclusively on the tensor amplitude, in contrast to the $\mathcal{L}_m$ coupling, which acts mainly on $n_s$. Even at $\alpha = 10^{8}$, the shift in $n_s$ is two orders of magnitude below the Planck uncertainty and $\Delta r \approx (4 -5)\times 10^{-4}$ is far below the current bound, so the placement in Fig.~\ref{fig:ns_r} is unaffected for all $\alpha$ considered. Larger $\alpha$ lies outside the regime of validity of the approximate
background equations and is not treated here. Its effect is comparatively small in the low-$\alpha$ regime explored by the
present analysis. We therefore use $\alpha=0$ in the analysis in order to isolate the phenomenology associated with the matter-torsion coupling $\beta$ without introducing an additional parameter degeneracy. 

\begin{table*}[t]
\centering
\caption{Dependence of the inflationary observables on the $T^2$ coupling
$\alpha$.}
\label{tab:alpha_dependence}
\begin{ruledtabular}
\begin{tabular}{ccccccc}
$N$ & $\alpha$ & $\phi_\star$ & $\epsilon_1$ & $\epsilon_2$
& $n_s$ & $r$ \\
\hline
50 & $0$       & 5.243968 & $2.621204\times10^{-4}$ & $3.789182\times10^{-2}$
   & 0.96158394 & $4.193926\times10^{-3}$ \\
50 & $10^{2}$  & 5.243968 & $2.621204\times10^{-4}$ & $3.789182\times10^{-2}$
   & 0.96158394 & $4.193927\times10^{-3}$ \\
50 & $10^{4}$  & 5.243962 & $2.621231\times10^{-4}$ & $3.789180\times10^{-2}$
   & 0.96158396 & $4.193970\times10^{-3}$ \\
50 & $10^{6}$  & 5.243341 & $2.623931\times10^{-4}$ & $3.788966\times10^{-2}$
   & 0.96158555 & $4.198289\times10^{-3}$ \\
50 & $10^{8}$  & 5.183190 & $2.952391\times10^{-4}$ & $3.765980\times10^{-2}$
   & 0.96174973 & $4.723825\times10^{-3}$ \\
\hline
55 & $0$       & 5.353418 & $2.187447\times10^{-4}$ & $3.456965\times10^{-2}$
   & 0.96499286 & $3.499916\times10^{-3}$ \\
55 & $10^{2}$  & 5.353418 & $2.187448\times10^{-4}$ & $3.456965\times10^{-2}$
   & 0.96499286 & $3.499916\times10^{-3}$ \\
55 & $10^{4}$  & 5.353411 & $2.187470\times10^{-4}$ & $3.456963\times10^{-2}$
   & 0.96499288 & $3.499953\times10^{-3}$ \\
55 & $10^{6}$  & 5.352784 & $2.189745\times10^{-4}$ & $3.456779\times10^{-2}$
   & 0.96499426 & $3.503592\times10^{-3}$ \\
55 & $10^{8}$  & 5.291966 & $2.466483\times10^{-4}$ & $3.436991\times10^{-2}$
   & 0.96513680 & $3.946372\times10^{-3}$ \\
\hline
60 & $0$       & 5.453801 & $1.853441\times10^{-4}$ & $3.178520\times10^{-2}$
   & 0.96784411 & $2.965505\times10^{-3}$ \\
60 & $10^{2}$  & 5.453801 & $1.853441\times10^{-4}$ & $3.178520\times10^{-2}$
   & 0.96784411 & $2.965506\times10^{-3}$ \\
60 & $10^{4}$  & 5.453794 & $1.853460\times10^{-4}$ & $3.178519\times10^{-2}$
   & 0.96784412 & $2.965537\times10^{-3}$ \\
60 & $10^{6}$  & 5.453161 & $1.855404\times10^{-4}$ & $3.178358\times10^{-2}$
   & 0.96784534 & $2.968646\times10^{-3}$ \\
60 & $10^{8}$  & 5.391764 & $2.091818\times10^{-4}$ & $3.161123\times10^{-2}$
   & 0.96797041 & $3.346909\times10^{-3}$ \\
\end{tabular}
\end{ruledtabular}
\end{table*}


\section{Concluding Remarks}\label{sec:conclsn}
In this work, we have investigated slow-roll inflation within the teleparallel framework of GR. Starting from the modified Friedmann and Klein-Gordon equations, we expressed the slow-roll parameters $\epsilon_1$ and $\epsilon_2$ entirely in terms of the Starobinsky potential. This allowed us to compute the scalar
spectral index $n_s$, the tensor-to-scalar ratio $r$, and the running
$\alpha_s$ for $N = 50$, $55$, and $60$ e-folds, and to compare the predictions with current CMB data.

In the limit $\beta \to 0$, $\alpha = 0$ the model reproduces the GR Starobinsky predictions, $n_s = 0.9616 -0.9678$ and
$r = (4.19 -2.96)\times 10^{-3}$ for $N = 50 -60$
(Table \ref{tab:GR_benchmark}), with $N = 55$ matching the Planck 2018 central
value $n_s = 0.9649$ to within $10^{-4}$.
The coupling $\beta$ affects the observables non-monotonically.
For $\beta \lesssim 10^{-6}$ its effect is negligible.
In the range $10^{-6} \lesssim \beta \lesssim 10^{-4}$, $n_s$ is enhanced by
up to $(1.3 -1.6)\times 10^{-3}$ at
$\beta \approx (2.4 -3.3)\times 10^{-5}$ and $r$ by $\approx 8\%$ at that
point, rising to $\approx (3.3 -4.7)\times 10^{-3}$ at its own maximum
near $\beta \approx 10^{-4}$ (Table \ref{tab:ns_maximum},
Figs. \ref{fig:ns_beta}-\ref{fig:r_beta}).
For $\beta \gtrsim 10^{-4}$ both quantities fall rapidly, and $n_s$ drops below
the Planck 95\% range for $\beta$ of a few $\times 10^{-4}$, which provides an
upper bound on the coupling from the inflationary epoch alone.

The $T^{2}$ coupling has a much weaker effect within the low-$\alpha$ regime
considered, $0 \le \alpha \le 10^{8}$ (Table \ref{tab:alpha_dependence}).
It is negligible for $\alpha \lesssim 10^{6}$ and at $\alpha = 10^{8}$ raises
$r$ by $\approx 13\%$ while shifting $n_s$ by only $\sim 10^{-4}$. Therefore, it
modifies the tensor amplitude without altering the compatibility with the
spectral-index constraints.

Within the allowed range, all predictions lie in the concave region of the
$(n_s, r)$ plane, inside the 95\% contour of the joint Planck+BK18+BAO
analysis for every $N$ between 50 and 60 and inside the 68\% contour for
$N \gtrsim 55$ (Fig. \ref{fig:ns_r}).
The Planck central value $n_s = 0.965$ is reached for $N = 55$ at
$\beta \approx 1.5\times 10^{-7}$ and $8.5\times 10^{-5}$ and for $N = 60$ at
$\beta \approx 1.4\times 10^{-4}$, but cannot be reached for $N = 50$
(Table \ref{tab:ns_constraint}).
Larger $N$ is therefore favored, independently of $\beta$.

The tensor-to-scalar ratio remains within
$r \approx (3.0 -4.7)\times 10^{-3}$ throughout, an order of magnitude
below the current bound $r(k=0.05) < 0.036$ from BICEP but at the sensitivity
projected for BICEP Array, LiteBIRD, and
CMB-S4.
The running is negative and small, decreasing in magnitude from
$\alpha_s = -(5.2 -7.5)\times 10^{-4}$ in the Starobinsky limit to
$-(4.0 -6.2)\times 10^{-4}$ at $\beta = 10^{-4}$
(Table \ref{tab:running}, Fig. \ref{fig:alphas_beta}), and is consistent with
the Planck results\cite{Planck:2018X}. 

The main conclusion is that small couplings ($\beta \lesssim 10^{-5}$, $\alpha \lesssim 10^{6}$)
reproduce Starobinsky inflation, while $\beta \gtrsim \mathcal{O}(10^{-4})$ is ruled out by $n_s$.
A precise determination of the upper bound on $\beta$, and of the preferred $N$, requires a treatment of reheating in this model.
Extending the analysis beyond the low-$\alpha$ regime, where the approximate
background equations cease to hold, and studying the influence of both
couplings on the post-inflationary and late-time dynamics, would further
constrain the parameter space.
Finally, a detection of $r \sim \mathcal{O}(10^{-3})$ by next-generation CMB polarisation experiments would provide a decisive test of the Starobinsky-like plateau predicted by this model.

\section*{Acknowledgment} SP acknowledges the Council of Scientific and Industrial Research (CSIR), Govt. of India, for providing Junior Research Fellowship (E-certificate No.: 24D/01/00653). PKS acknowledges IUCAA, Pune, India for providing support through the visiting Associateship program. 


\appendix

\bibliographystyle{unsrtnat}
\bibliography{biblio}

\end{document}